# Nanoscale SOI-MOSFETs with Electrically Induced Source/Drain Extension: Novel attributes and Design considerations for Suppressed Short-channel Effects

Ali A. Orouji[1] and M. Jagadesh Kumar[2*]

[1]Department of Electrical Engineering, Semnan University, Semnan, IRAN

[2]Department of Electrical Engineering, Indian Institute of Technology, Delhi, New Delhi, INDIA

[*]Corresponding author, Email: mamidala@ieee.org, Fax: 91-11-2658 1264

***Abstract***- Design considerations for a below 100 nm channel length SOI MOSFET with electrically induced shallow source/drain junctions are presented. Our simulation results demonstrate that the application of induced source/drain extensions to the SOI MOSFET will successfully control the SCEs and improve the breakdown voltage even for channel lengths less than 50 nm. We conclude that if the side gate length equals the main gate length, the hot electron effect diminishes optimally.

## I. Introduction

Silicon-On-Insulator (SOI) has been the forerunner of the CMOS technology in the last decade offering superior CMOS devices with higher speed, higher density, excellent radiation hardness and second order effects for submicron VLSI applications [1]. But, these devices suffer from the hot electron effect and the short-channel effects such as steep threshold voltage roll-off due to charge-sharing between drain/source and channel and increased off-state leakage current due to sensitivity of the source/channel barrier to the drain potential or drain-induced barrier lowering (DIBL). Therefore, reduction of hot electron and short cannel effects (SCEs) plays a significant role in down-scaling the CMOS technology. Various approaches have been proposed to control these problems [2-8].

One of the effective methods to control SCEs is the use of an ultra shallow extended source/drain. But, it is very difficult to form shallow junctions by conventional fabrication techniques. There have been many reports that explained suppression of SCEs using an inversion layer as an ultra shallow extended source/drain [9]-[15] which can be created employing a side gate on both sides of the main gate and this device is known as the EJ-SOI MOSFET in literature. In spite of the area penalty associated with these devices due to the additional side gates required for the inversion layer formation, EJ MOSFETs are expected to play a major role in the reduction of SCEs and hot carrier effects. In this paper, we discuss the device design for sub 100 nm channel lengths with a view to improve the performance of the device in terms of the hot electron and short-channel effects. We demonstrate that if the side gate length equals the main gate length, the device exhibits the least hot electron effect. Also, our results demonstrate the unique attributes of the EJ-SOI MOSFET structure in suppressing the hot electron effect and SCEs for channel lengths even less than 50 nm.

## II. EJ-SOI MOSFET Structure

A schematic cross-sectional view of the EJ-SOI MOSFET implemented using the 2-D device simulator MEDICI [16] is shown in Fig.1. As can be seen from the figure, extremely shallow junctions are realized in SOI MOSFETs with a triple-gate structure consisting of one main gate and two side gates. The two side gates are biased independently of the main gate, and induce the inversion layers that work as the virtual source or drain. The lengths of the main and side gates are $L_M$ and $L_S$, respectively. Silicon dioxide is adopted as a diffusion barrier between the main gate and the side gate. The doping in the silicon thin layer is kept at $6\times10^{16}$ $cm^{-3}$ [17,18]. The device parameters used in our simulation are shown Table I. It is worth noting that the thickness of the diffusion barriers ($t_d$ = 2 nm) is chosen to be significantly smaller than the main and the side gate lengths. It cannot be smaller than the gate oxide thickness. If we choose a thicker $t_d$, there will be a discontinuity between the main and side gate regions in the channel. All the device parameters of the EJ-SOI MOSFET are equivalent to those of the Conventional SOI (C-SOI) MOSFET unless otherwise stated.

It is worth noting that we have included the full energy balance model in our simulations. This is because the thermal equilibrium approximation (TEA) is invalid in devices like EJ-SOI MOSFET. The TEA approximation implies that carrier temperatures are equal to the (constant) lattice temperature. In some devices, electric fields in the direction of current flow can be so large that the carrier velocity in the whole channel is in saturation and consequently, carrier temperatures must differ significantly from the lattice temperature in the whole channel region of the device [17-21].

## III. Results and Discussions

Fig. 2 shows the typical MEDICI simulated surface potential profiles in the channel for the main channel length of $L_M$=50 *nm* and the side channel length of $L_S$=50*nm* for a side gate voltage $V_{SGS}$=1.5 V and main gate voltage $V_{MGS}$= 0 V. It can be seen from the figure that due to the presence of the electrically shallow junction (EJ) in the EJ-SOI MOSFET, there is no significant change in the surface potential under the main gate as the drain bias is increased even for drain voltages up to 1.5 V. Hence, the channel region under the main gate is “screened” from the changes in the drain potential, i.e. the drain voltage is not absorbed under the main gate. As a consequence, we expect that variation of threshold voltage will be small and under control. In Fig. 3, the threshold voltage of EJ-SOI structure as a function of the main channel length is compared with the C-SOI MOSFET. As mentioned above, it can be observed clearly that as the channel length is decreased, the EJ-SOI structure exhibits lower threshold voltage roll-off than the C-SOI structure.

In Fig. 4, the electric field distribution along the channel is shown for the C-SOI and the EJ-SOI MOSFETs. It is evident from the figure that the presence of an extremely shallow inversion layer reduces the peak electric field considerably. With reduction in the peak electric field, the breakdown voltage of the device should improve. The output characteristics of the EJ-SOI structure are shown in Fig. 5. As can be seen from the figure, the breakdown voltage and the drain conductance of the EJ-SOI structure improve when compared to the C-SOI structure. The reduction in the current of EJ-SOI device is due to the series resistance of the side gate regions.

## IV. Design issues of EJ-SOI MOSFET

The side gate bias and side gate lengths are two important problems in the design of an EJ-SOI MOSFET. Fig. 6 shows the dependence of the surface potential on the side gate bias when the main gate voltage is set at 1.5 V. It is clear from Fig. 3, that the threshold voltage of device is 0.52 V at $V_{SGS}$ =1.5 V. Therefore, the region under the main gate is in strong inversion when 1.5 V is applied to the main gate. When the side gate voltage is less than 0.17 V (threshold voltage of the side gates is 0.17 V), the surface-potential in the region under the side gate or virtual source/drain regions is low, and therefore, there are not enough carriers in the virtual source/drain regions. Contrarily, when the side gate bias is more than 0.17 V, the potential in the virtual source/drain will increase and the carrier concentrations in the inversion layers are sufficiently large to function as the virtual source/drain regions. If the side gate voltage is increased, the resistance of the virtual source /drain regions will also reduce and the drain current will increase. It should be noted that the potential edges in the virtual source/drain region move slightly toward the channel as the side gate bias increases beyond 0.17 V. This indicates that the channel length depends only slightly on the side gate bias.

For investigating the side gate length effects in the EJ-SOI structure, a comparison in terms of threshold voltage, vertical electric field and electron temperature is performed. The threshold voltage versus the side gate lengths is shown in Fig. 7. It is evident from the figure that the threshold voltage of the device decreases with side gate length reduction. Because, the movement of potential edges in the virtual source/drain regions toward the main channel are comparable with the main channel length in small side gate lengths.

Fig. 8 (a) and 8 (b) show the electron temperature at the surface of the silicon thin layer for EJ-SOI and C-SOI structures with fixed main gate lengths of 50 nm and 40 nm, respectively. As can be seen from the figures, the electron temperature will increase with reduced side gate length.  However, we notice that the hot carrier effect will be significantly lower in the EJ-SOI structure as compared to the C-SOI structure whenever the side gate length is equal the main gate length.

Fig. 9 shows the electric field in vertical direction at 5 nm from the edge of the side gate region on the drain side with a fixed main gate length of 50 nm. Two conclusions can be drawn from the figure. First, it can be seen that the peak electric field at the back gate side is large compared to the front gate side. Second, the electric field increases with decreasing the side gate length. Fig. 10 (a) and 10 (b) show the electron temperature at the same position that the electric field is drawn in Fig. 9 with fixed main gate lengths of 50 nm and 40 nm, respectively. It can be seen from the figures that due to the increase in vertical electric field with reduction of side gate length, the electron temperature will increase. In conclusion, we observe that due to the high electric field near drain side, the electron temperature is different from the lattice temperature and the side gate length plays an important role in the reduction of the hot carrier effects in the device.

## V. Conclusion

Design and performance considerations of a SOI MOSFET with electrically shallow junction as source/drain have been studied using two dimensional simulation. The EJ-SOI MOSFET device has a triple-gate structure consisting of one main gate and two side gates. The two side gates are biased independently of the main gate, and induce the inversion layers that work as the virtual source or drain. Our results establish that the EJ structure, exhibits subdued SCEs even for channel lengths far below 100 nm due to a very thin inversion layer in the induced source and drain extensions. Also, the hot electron effects with increasing the side gate length can be more effectively reduced in the EJ structure. Further, it is clearly seen that the EJ structure gives rise to improved breakdown voltage as well as output conductance. Thus, the EJ structure opens up a new avenue to improve the hot electron and short-channel behaviors of the SOI MOSFETs over their single-gate SOI and the bulk counterparts for channel lengths below 100 nm.

## Figure Captions

Figure 1 Cross-sectional view of the EJ-SOI MOSFET

Figure 2 Surface potential profiles in the channel of EJ-SOI MOSFET and C-SOI MOSFET structures.

Figure 3 Threshold voltage versus main channel length for channel lengths up to 20 nm with $V_{DS} = 50$ mV.

Figure 4 Longitudinal electric field in the channel of EJ-SOI MOSFET and C-SOI structures.

Figure 5 Output characteristics of EJ-SOI MOSFET and C-SOI MOSFET structures.

Figure 6 Surface potential in the channel of EJ-SOI MOSFET for different side gate voltages for $V_{MGS} = 1.5$ V and $V_{DS} = 0$ V.

Figure 7 Threshold voltage versus side gate length for side gate lengths up to 20 *nm* for $V_{DS} = 50$ mV and $V_{SGS} = 1.5$ V.

Figure 8 Electron temperature at the surface of silicon thin layer in the EJ-SOI MOSFET and C-SOI MOSFET structures for a) $L_M = 50$ nm and b) $L_M = 40$ nm.

Figure 9 Vertical electric field profiles at 5 nm from the edge of the side gate region on the drain side.

Figure 10 Electron temperature profiles in vertical direction at 5 nm from the edge of the side gate region on the drain side for a) $L_M = 50$ nm and b) $L_M = 40$ nm.

Table I: Simulation parameters

| Parameter | Value |
|---|---|
| Silicon thin layer doping | $6\times10^{16}$ $cm^{-3}$ |
| Source/drain doping | $5\times10^{19}$ $cm^{-3}$ |
| Work function of the side gates | 4.7 eV |
| Work function of the main gate | 4.9 eV |
| Silicon thin layer thickness | 50 nm |
| Buried oxide thickness | 500 nm |
| Gate oxide thickness | 2 nm |
| Barrier diffusion thickness | 2 nm |
| Typical main gate length | 50 nm |
| Typical side gate length | 50 nm |

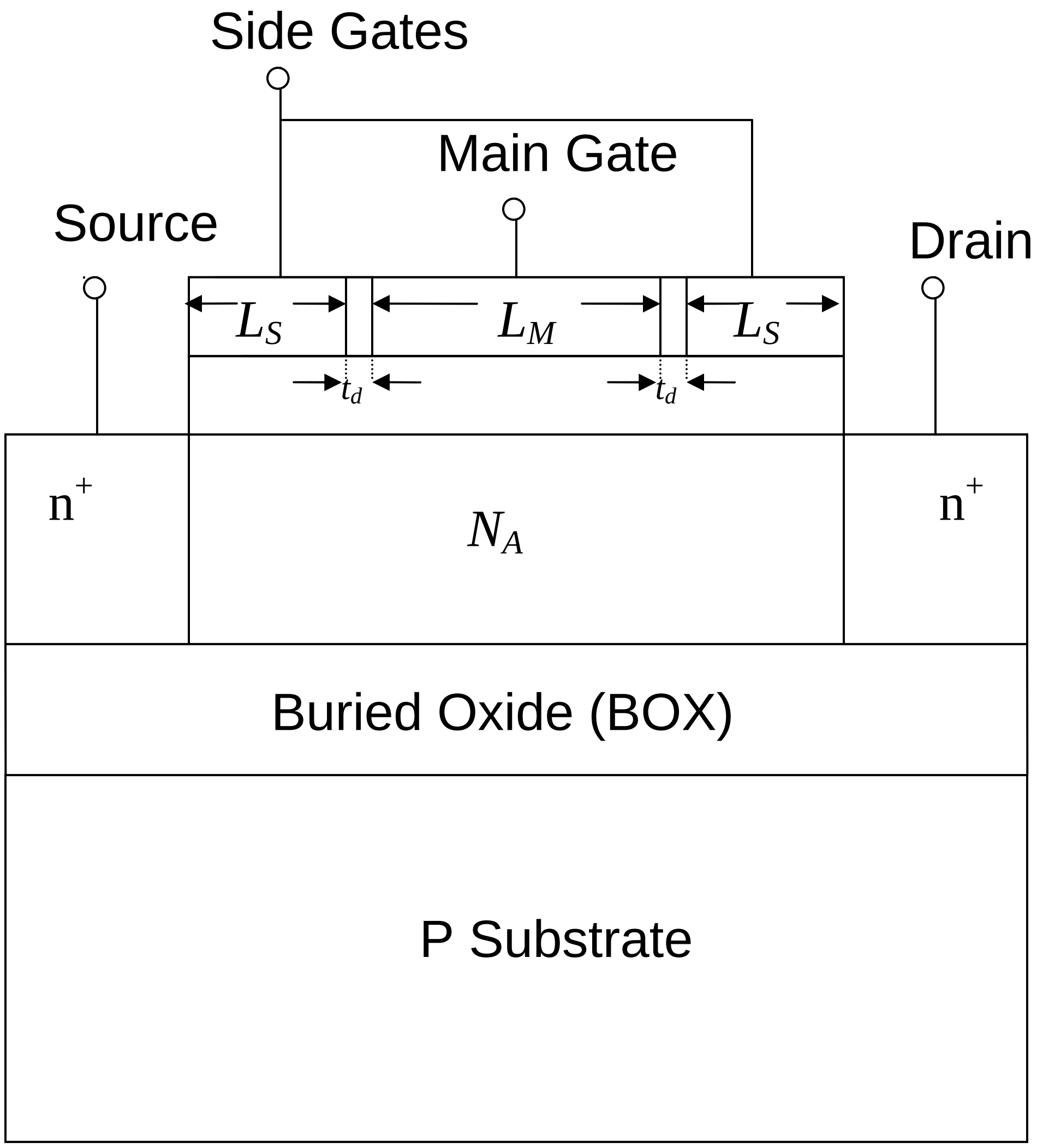

Fig. 1.

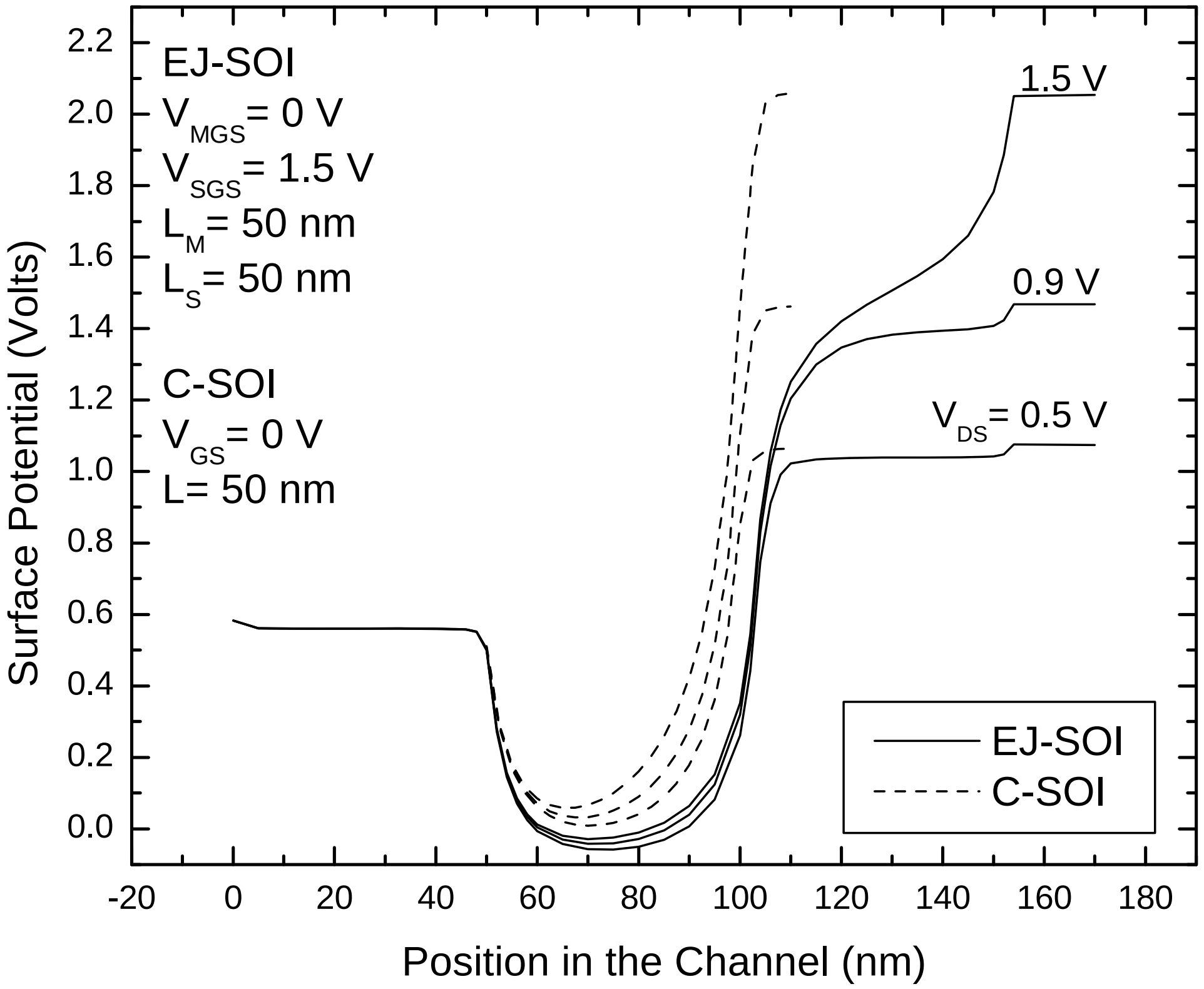

Fig. 2.

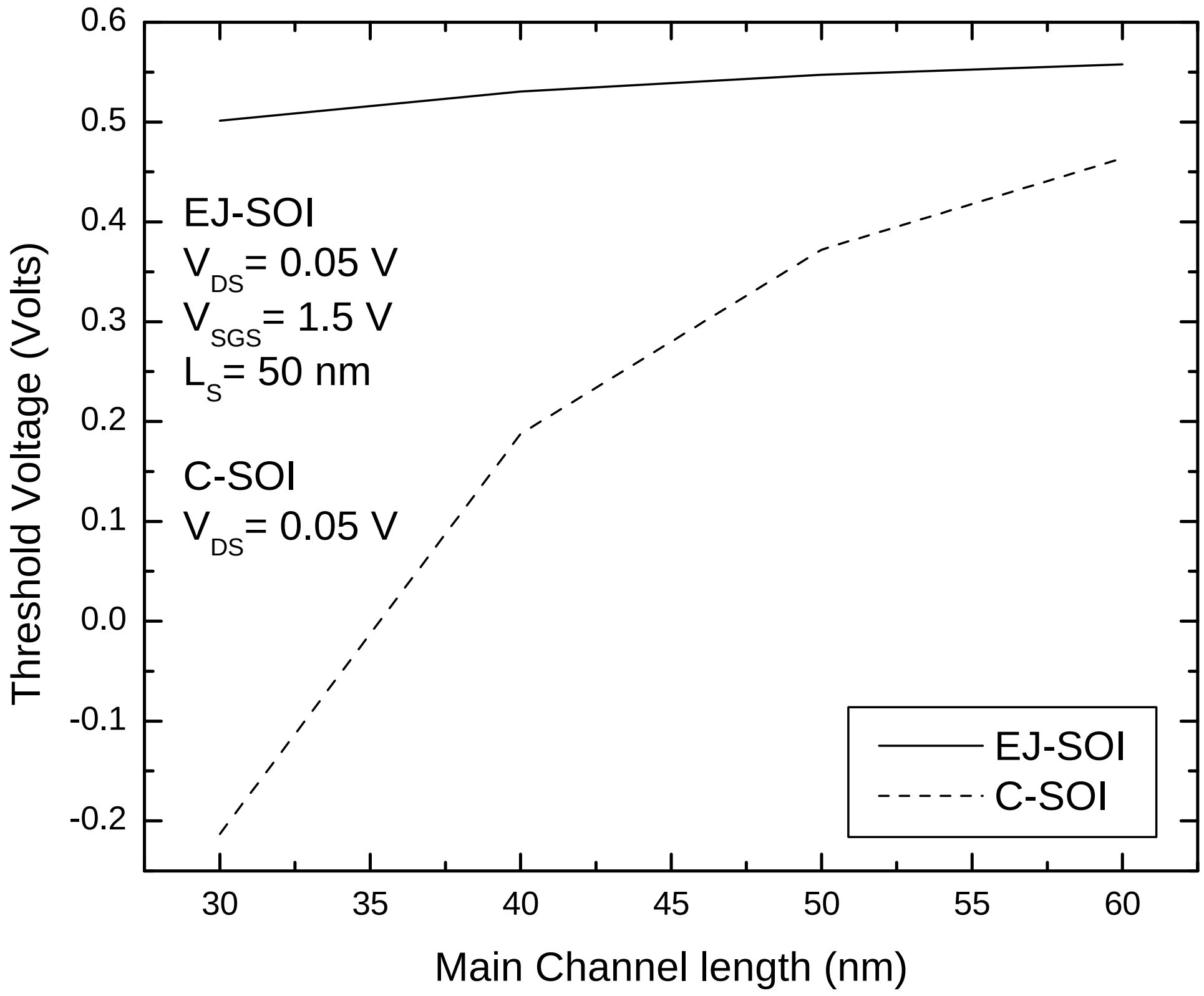

Fig. 3.

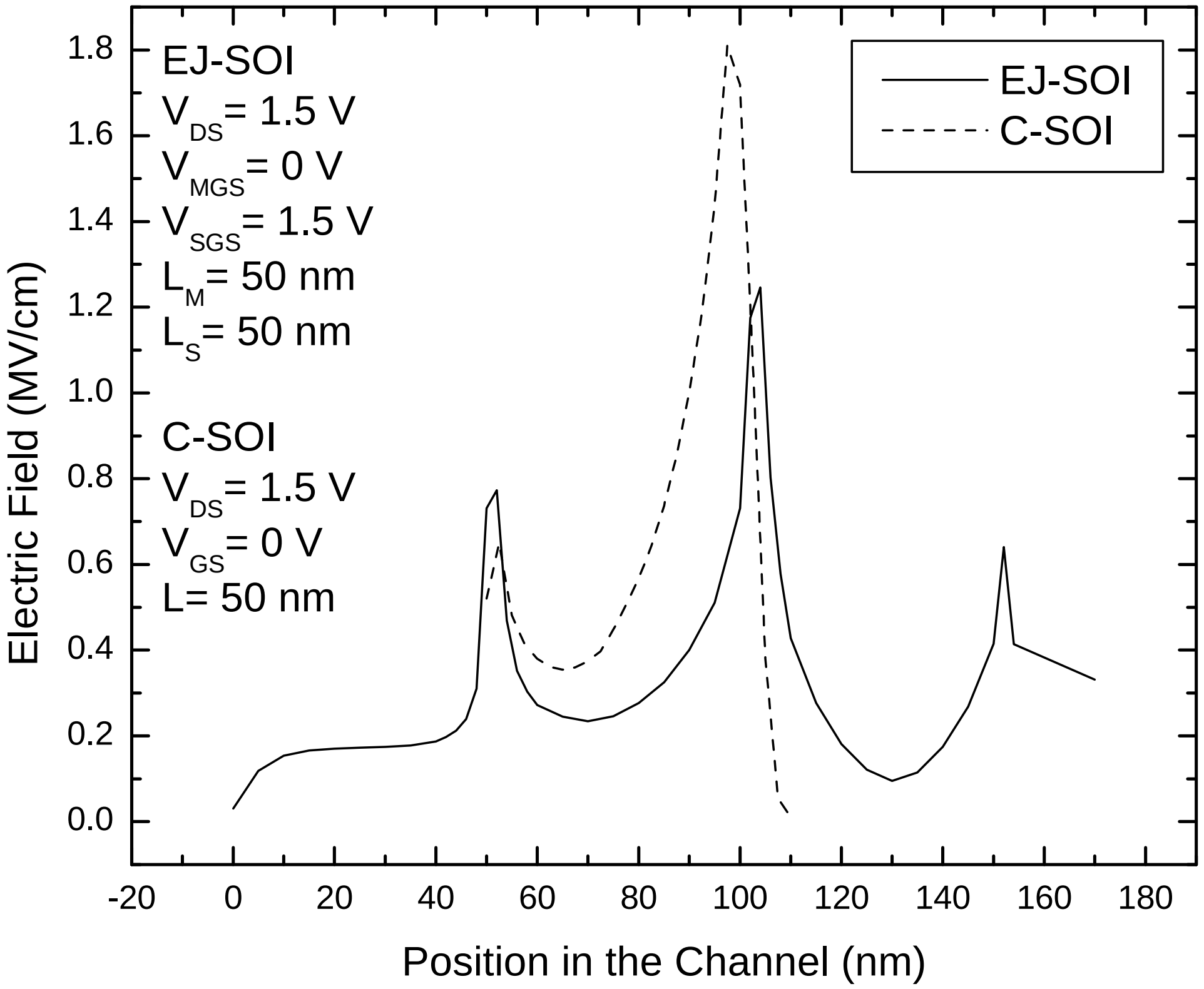

Fig. 4.

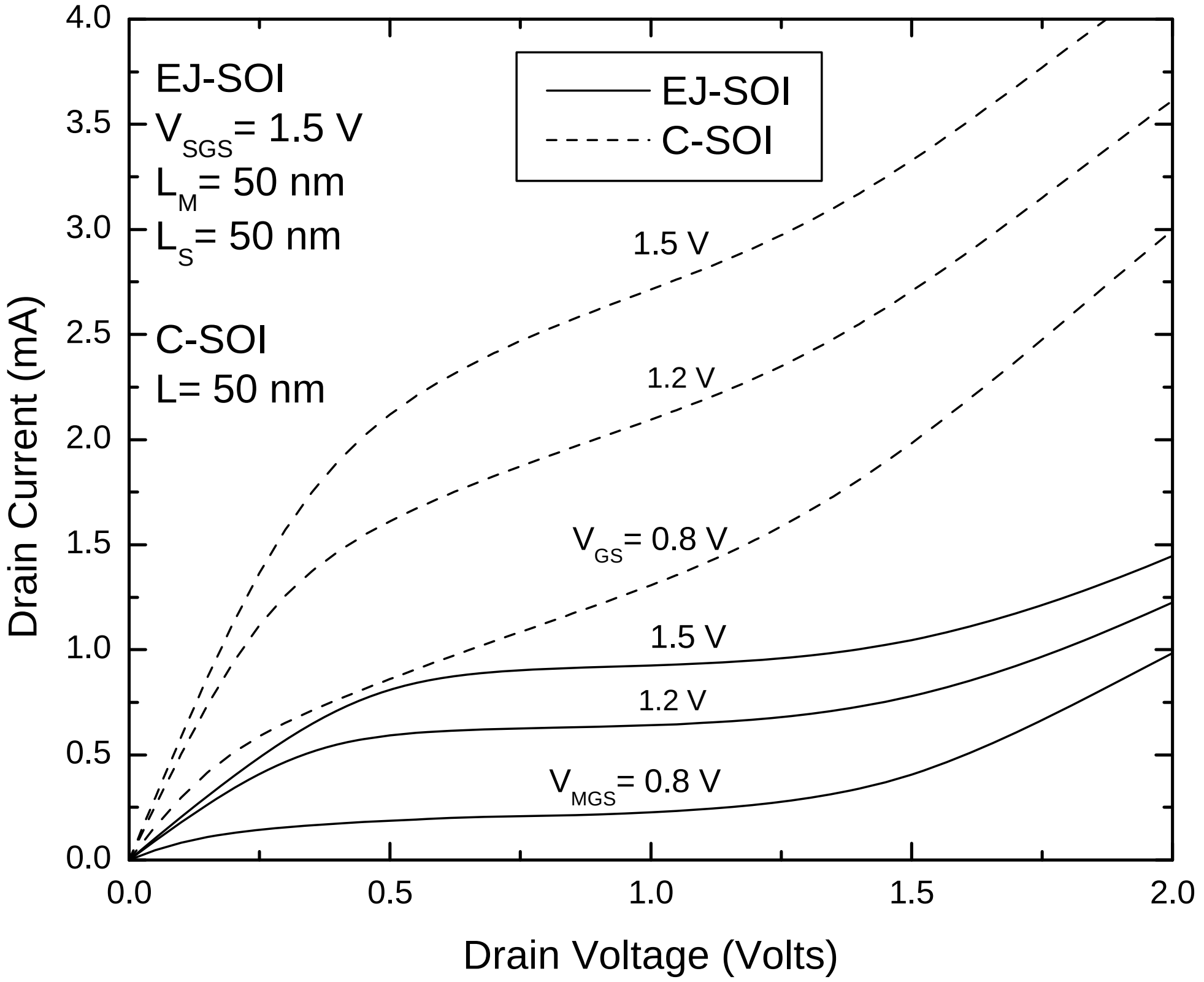

Fig. 5.

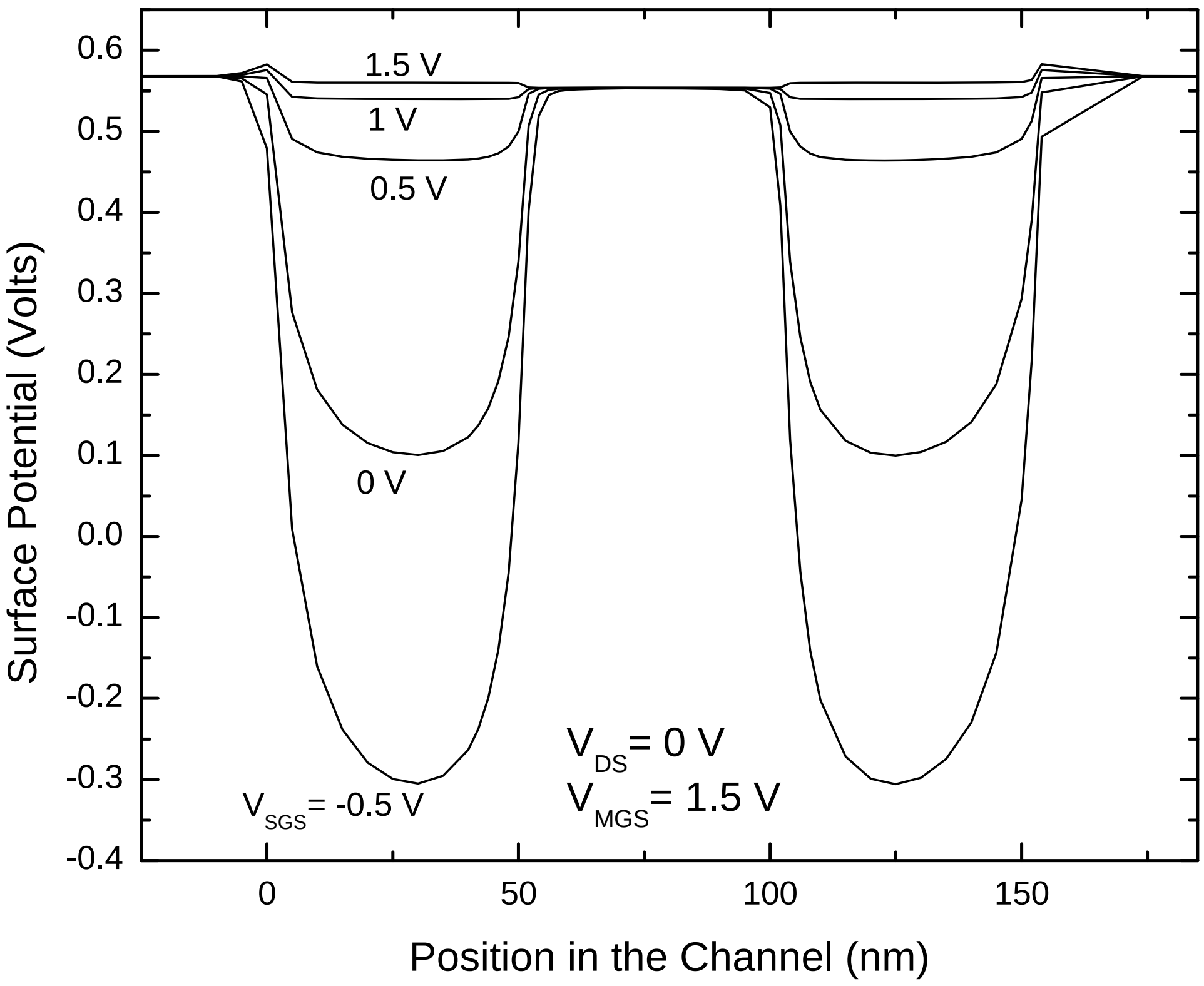

Fig. 6.

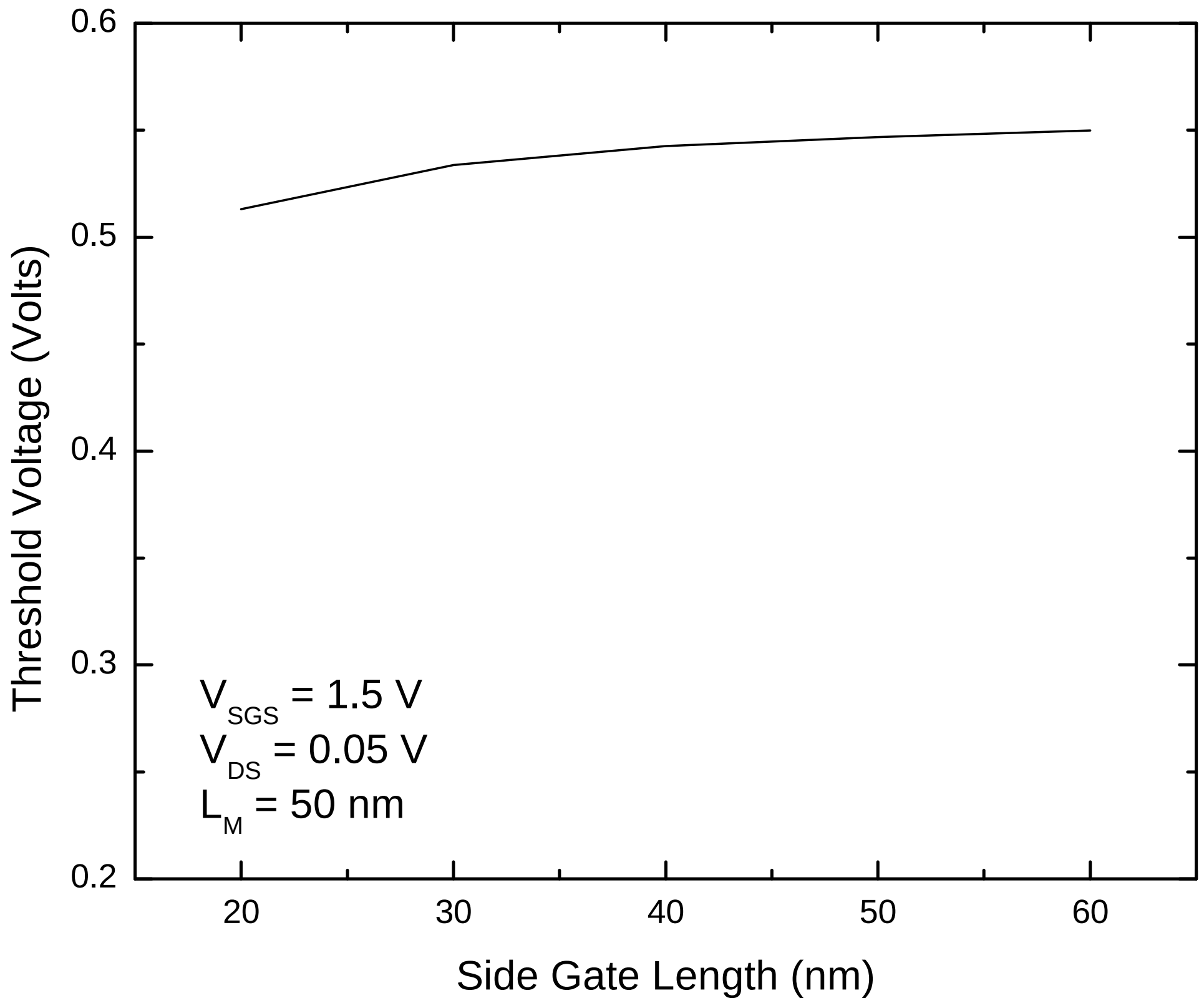

0.6
0.5
0.4
0.3
0.2
Threshold Voltage (Volts)
$V_{SGS}$ = 1.5 V
$V_{DS}$ = 0.05 V
$L_M$ = 50 nm
20
30
40
50
60
Side Gate Length (nm)

Fig. 7.

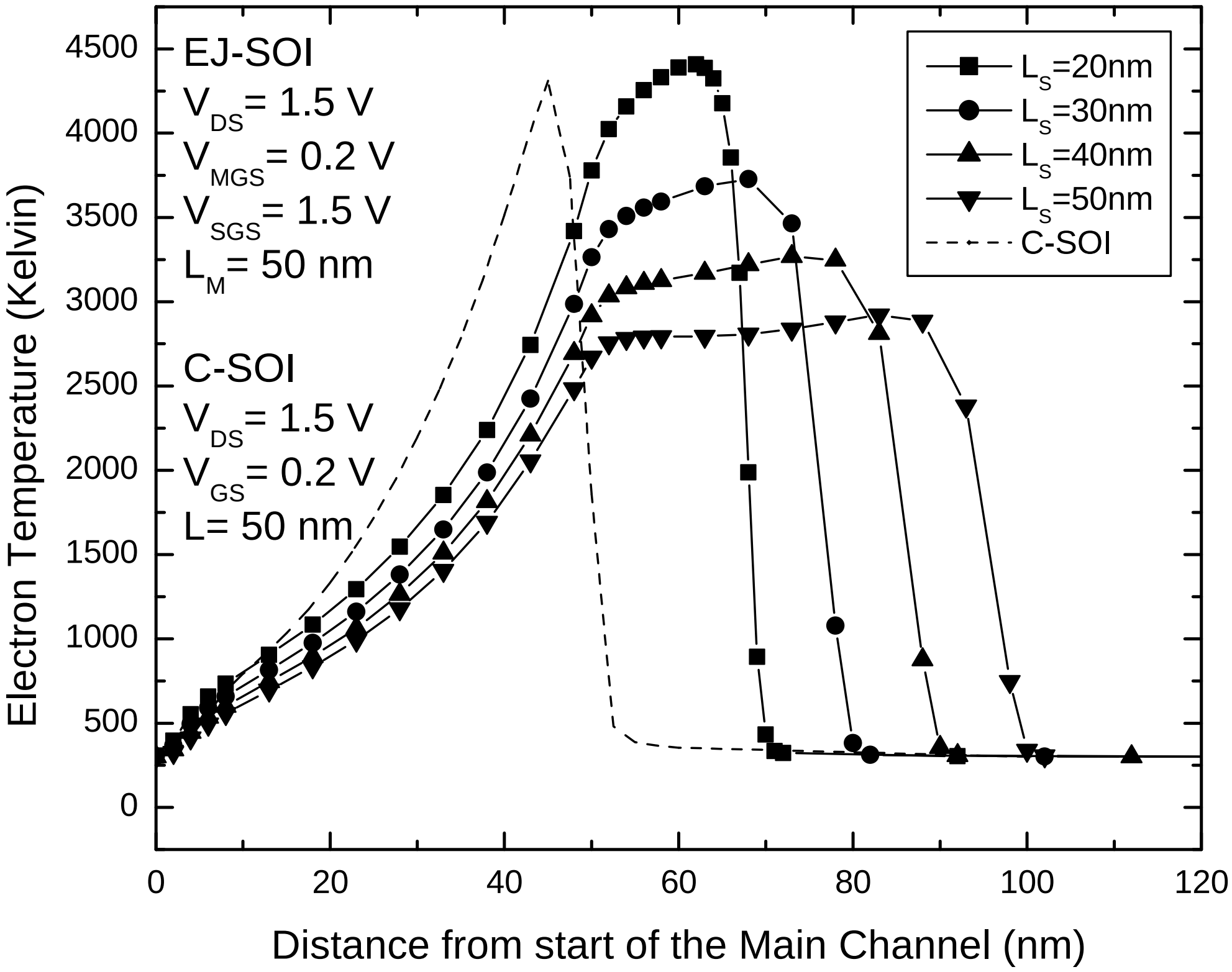
EJ-SOI
$V_{DS}$= 1.5 V
$V_{MGS}$= 0.2 V
$V_{SGS}$= 1.5 V
$L_M$= 50 nm
C-SOI
$V_{DS}$= 1.5 V
$V_{GS}$= 0.2 V
L= 50 nm
$L_S$=20nm
$L_S$=30nm
$L_S$=40nm
$L_S$=50nm
C-SOI
Electron Temperature (Kelvin)
Distance from start of the Main Channel (nm)
0
500
1000
1500
2000
2500
3000
3500
4000
4500
0
20
40
60
80
100
120

Fig. 8 (a)

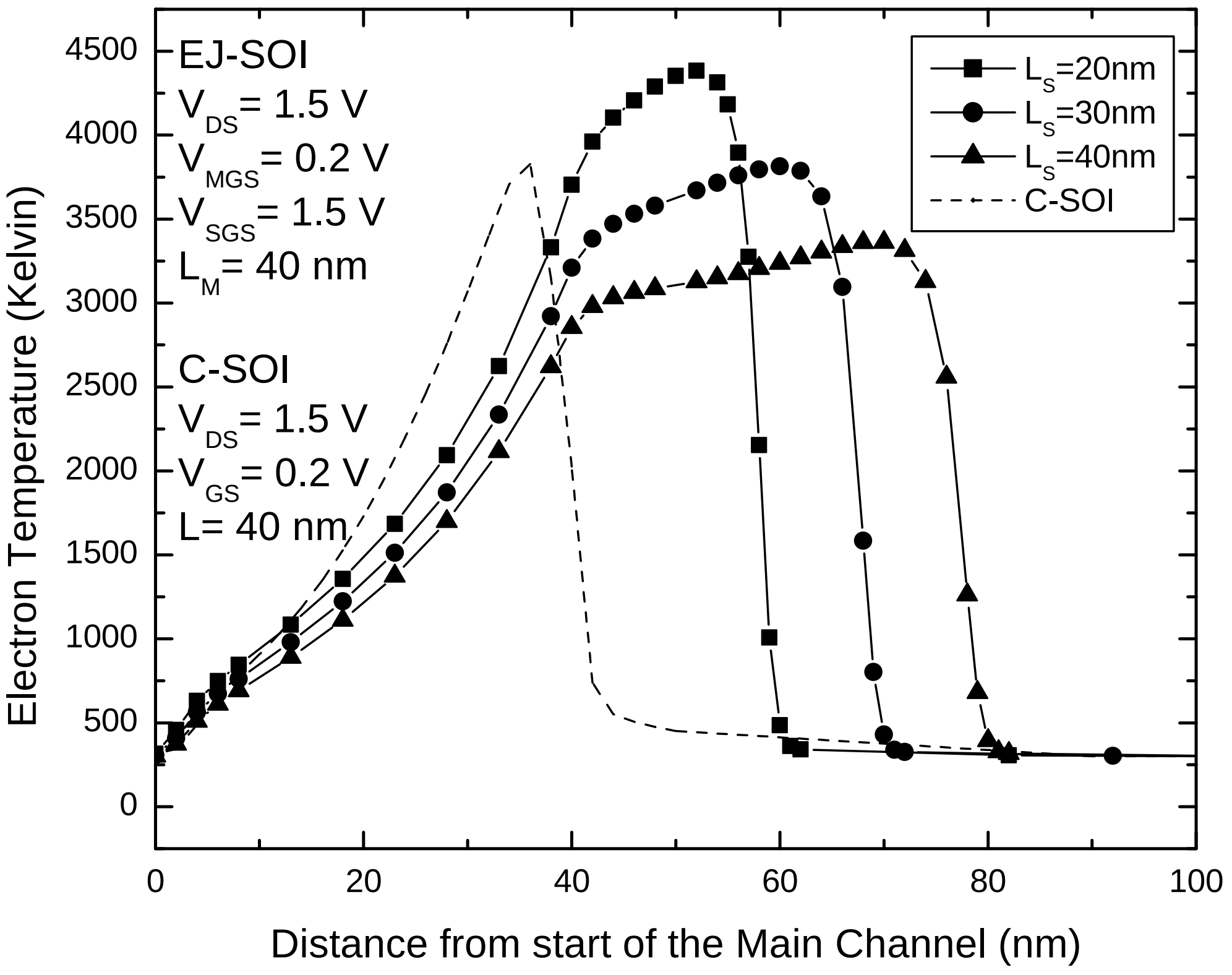
EJ-SOI
VDS= 1.5 V
VMGS= 0.2 V
VSGS= 1.5 V
LM= 40 nm
C-SOI
VDS= 1.5 V
VGS= 0.2 V
L= 40 nm
LS=20nm
LS=30nm
LS=40nm
C-SOI
Electron Temperature (Kelvin)
Distance from start of the Main Channel (nm)
0
500
1000
1500
2000
2500
3000
3500
4000
4500
0
20
40
60
80
100

Fig. 8 (b)

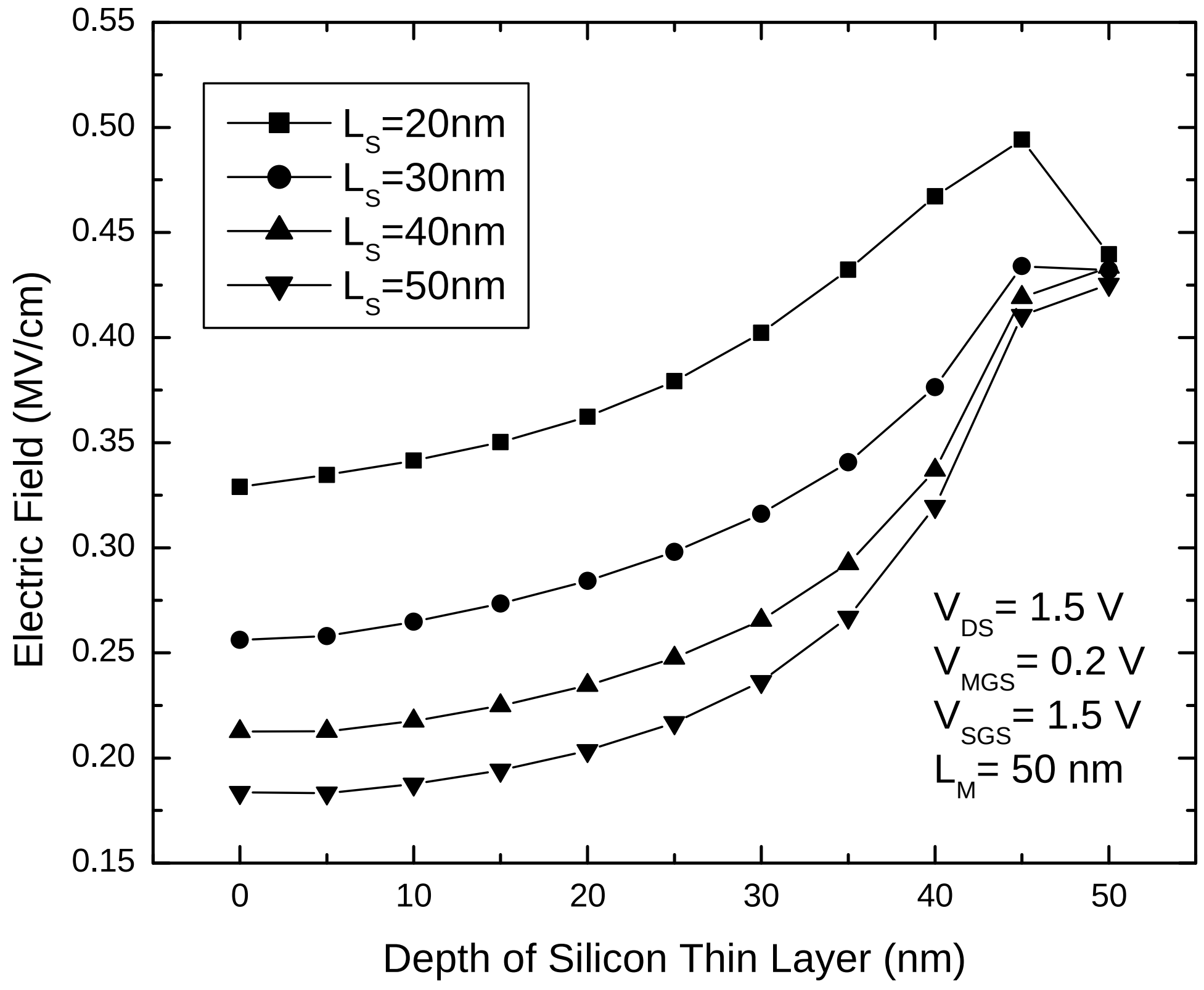

Fig. 9.

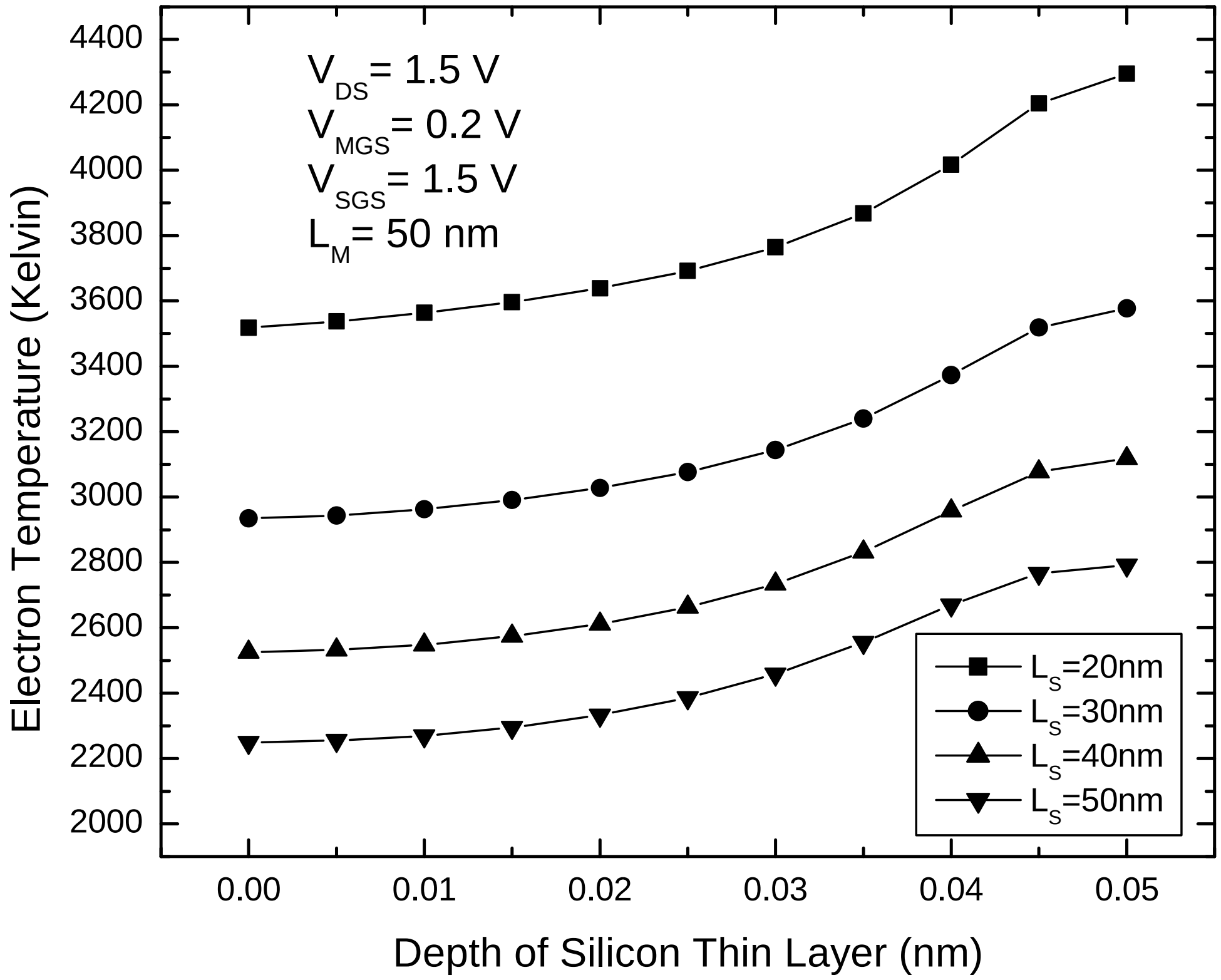

Fig. 10 (a)

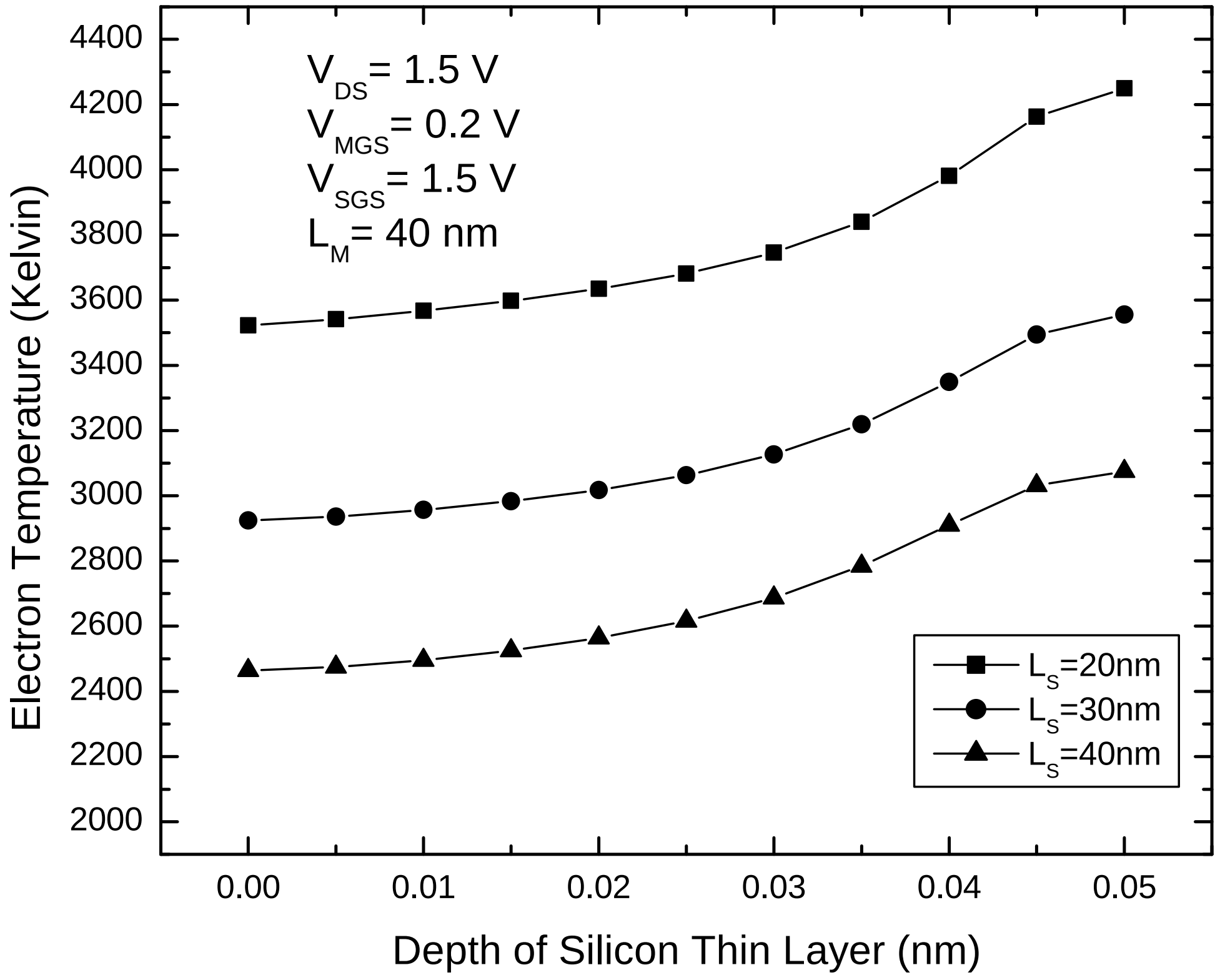

Fig. 10 (b)